\documentclass[12pt]{article}

\def\bn{\bigskip\noindent}

\usepackage{color}
\usepackage[dvips]{graphicx}
\usepackage{amsmath,amssymb}

\begin{document}
\begin{titlepage}
\renewcommand{\thefootnote}{\fnsymbol{footnote}}
 \font\csc=cmcsc10 scaled\magstep1
 {\baselineskip=14pt
 \rightline{
 \vbox{\hbox{November, 2003}
       \hbox{UT/03-36} 
       \hbox{hep-th/0311220}
      }}}
\bn
\begin{center}
\Large{ \bf Asymptotic Form of  Gopakumar--Vafa Invariants
\\
from Instanton Counting
}
\vspace*{1.5cm}

\normalsize{Yukiko Konishi${^1}$ and
            Kazuhiro Sakai${^2}$}

\vspace*{1cm}

\normalsize{\it ${^1}$
Research Institute for Mathematical Sciences, \\
Kyoto University, \\
Kyoto 606-8502, Japan}\\
\normalsize{\texttt{konishi@kurims.kyoto-u.ac.jp}}

\normalsize{\it ${^2}$
Department of Physics, Faculty of Science,\\
University of Tokyo,\\
Hongo 7-3-1, Bunkyo-ku, Tokyo 113-0033, Japan}\\
\normalsize{\texttt{sakai@hep-th.phys.s.u-tokyo.ac.jp}}

\vspace{2.5ex}

\end{center}
\vspace{1cm}
\begin{abstract}
We study the asymptotic form of the 
Gopakumar--Vafa invariants at all genera for 
Calabi--Yau toric threefolds which have the
structure of fibration of
the $A_n$ singularity  over ${\bf P}^1$.
We claim that the asymptotic form is 
the inverse Laplace transform of the 
corresponding instanton amplitude in the 
prepotential of ${\mathcal N}=2$ $SU(n+1)$
gauge theory coupled to external graviphoton
fields,
which is given by the
logarithm of the 
Nekrasov's partition function.
\end{abstract}
\end{titlepage}

\newcommand{\vsp}{\vspace*{1cm}}
\newcommand{\midasi}[1]{\noindent{\bf #1}}
\newcommand{\remarks}{\noindent{\bf Remarks}}
\newcommand{\remark}{\noindent{\bf Remark}}
\newcommand{\notation}{\noindent{\bf Notations}}
\newcommand{\w}[1]{l(#1)}
\newcommand{\len}[1]{d(#1)}
\newcommand{\conj}[1]{{#1}^{t}}
\newcommand{\R}{R}
\newcommand{\Q}{S}

\newcommand{\va}{\boldsymbol{a}}
\newcommand{\vt}{\boldsymbol{t}}
\newcommand{\vd}{\boldsymbol{d}}


\section{Introduction}
\hspace*{2.5ex}
In this article
we study the geometric engineering of the 
four-dimensional ${\mathcal N}=2$ $SU(n+1)$
gauge theory without matter hypermultiplets.
It is known that the gauge theory can be realized by
compactifying the type IIA string on a certain 
Calabi--Yau threefold and taking a certain limit 
\cite{KaKlVa}.
Such a Calabi--Yau threefold must satisfy two
conditions:
it must have
$A_n$-singularity so that the 
gauge group is $SU(n+1)$; it must be
a fibration over the Riemann surface of 
genus zero (i.e. ${\bf P}^1$)
so that the gauge theory has 
the asymptotic freedom.
The limit is a double scaling limit
such that the 
exceptional curves of the $A_n$ singularity shrink
and the base ${\bf P}^1$ expands simultaneously.
In this limit, 
the free energy of topological strings on the Calabi--Yau threefold
becomes 
the prepotential of the gauge theory.
The implication of this  
phenomenon is that
the worldsheet instanton correction
should reproduce the
spacetime instanton correction
in the gauge theory and the precise
relation between them is what we derive in this article.

In the language of mathematics 
we study the asymptotic form of the 
Gopakumar--Vafa invariants at all genera
for  Calabi--Yau threefolds
which are smooth
toric varieties
and possess the structure of the fibration of
$A_n$-singularity over ${\bf P}^1$.
Recently, Iqbal and Kashani-Poor showed that the
topological string amplitude 
obtained by the method of the geometric transition 
\cite{GoVa1, AgKlMaVa, AgMaVa}
agrees with Nekrasov's partition function for
instanton counting \cite{IqKa1,IqKa2}
in the limit which they call {\it field theory limit}.
This limit involves the limit of the string coupling
(the genus expansion parameter) as well as the usual
limit of the K\"ahler parameters.
By taking the logarithm,
one can see that 
the generating function of the Gromov--Witten invariants
of the Calabi--Yau threefold
agrees with the logarithm of 
Nekrasov's partition function \cite{Nek} 
in this limit: 
the genus zero part of the generating function
corresponds to the Seiberg--Witten prepotential \cite{SeWi1}
and the higher genus parts correspond to 
the effects due to the external graviphoton fields
since the genus expansion of the former
matches the expansion in terms of
$\hbar$ in the latter 
in the field theory limit.
From this relation,
we derive the asymptotic form of 
the Gopakumar--Vafa invariants
of the $A_n$-fibration over ${\bf P}^1$ at all genera:
it is  obtained as the inverse Laplace transform of the 
corresponding term in the logarithm of 
Nekrasov's partition function for $SU(n+1)$ instanton counting.

The asymptotic form of the
Gopakumar--Vafa invariants 
was first studied in 
the quintic case \cite{CaDeGrPa}.
Other cases studied so far are:
the canonical bundle of ${\bf P}^2$ and 
other one modulus local mirror systems with 
Picard-Fuchs equations given by Meijer's equation 
\cite{KlZa};
the canonical bundle of Hirzebruch surface 
${\bf F}_0,{\bf F}_1,{\bf F}_2$ \cite{KaKlVa};
the canonical bundle of ${\bf F}_2$  blown up 
at 1,2,3-points \cite{KoNa}.
The last two cases are the results of the 
geometric engineering of $SU(2)$ gauge theory.
All of these results concern
Gopakumar--Vafa invariants at genus zero. 
To our knowledge, our result is the first for
the higher genus cases.

The organization of the paper is as follows.
In section \ref{nekrasov}
we review  Nekrasov's partition function \cite{Nek, NaYo}.
Section \ref{toric} is devoted to
the description of the  $A_n$-fibration over ${\bf P}^1$.
In section \ref{iqka}
we review 
the results of Iqbal and Kashani-Poor \cite{IqKa1,IqKa2}.
In section \ref{gv} we derive the 
asymptotic form of the Gopakumar--Vafa invariants.
In
section \ref{exa2} and in appendix \ref{data},
we will test our results 
in the case of $A_2$-fibration over ${\bf P}^1$.
The data of local B-model is included in appendix \ref{localB}.

\section{Nekrasov's Partition Function}
\label{nekrasov}
\hspace*{2.5ex}
The closed formula for the Seiberg--Witten prepotential
of the  ${\mathcal N}=2$ $SU(n+1)$
gauge theory was derived by Nekrasov \cite{Nek}
and its mathematical proof was given  by Nakajima--Yoshioka
\cite{NaYo}.
The instanton correction part of the
prepotential is given by
\begin{equation}
-\hbar^2\log {\mathcal Z}_{{\rm Nekrasov}}^{A_n}|_{\hbar=0}.
\end{equation}
Here ${\mathcal Z}_{{\rm Nekrasov}}^{A_n}$ is 
Nekrasov's partition function for instanton counting.
The terms with higher order in $\hbar$  correspond to  the
effect  of the external graviphoton fields.
Before giving the form of ${\mathcal Z}_{{\rm Nekrasov}}^{A_n}$, 
let us  explain notations. 
We use the letter
$\R$, $\R_i$ for a partition (or a Young diagram)
and 
$\mu_i$, $\mu_{i,j}$ for its parts:
$\R=(\mu_1,\mu_2,\ldots)$, $\R_i=(\mu_{i,1},\mu_{i,2},\ldots)$.
$\w{\R}$ denotes 
the weight of $\R$ (the number of boxes of the Young diagram)
and 
$\len{\R}$ the length of $\R$ 
(the number of rows of the Young diagram).
$\conj{\R}$ denotes the conjugate partition of $\R$ 
(the transposed Young diagram).
$\kappa(\R):=\sum_{i=1}^{\len{\R}}\mu_i(\mu_i-2i+1)$.

Nekrasov's partition function is 
\begin{equation}
\label{nek}
{\mathcal Z}_{{\rm Nekrasov}}^{A_n}=
\sum_{\R_1,\ldots,\R_{n+1}}
\Lambda^{\w{\R_1}+\cdots+\w{\R_{n+1}}}
\prod_{i,j=1}^{n+1}\prod_{k,l=1}^{\infty}
\frac{a_{i,j}+\hbar(\mu_{i,k}-\mu_{j,l}+l-k)}{a_{i,j}+\hbar(l-k)}.
\end{equation}
The summation is over $n+1$ partitions
$\R_1,\ldots,\R_{n+1}$. 
$\Lambda$ is the dynamical scale.
$a_{i,j}=a_i-a_j$, where
$a_1,\ldots,a_{n+1}$ are the vacuum expectation values of the 
complex scalar fields in the gauge multiplet
of unbroken $U(1)^n\subset SU(n+1)$
of the ${\mathcal N}=2$ 
$SU(n+1)$ gauge theory.
From the mathematical viewpoint,
Nekrasov's partition function is 
the integration in the equivariant cohomology over the 
moduli space of instantons on ${\bf R}^4$
and $a_1,\ldots,a_{n+1}$ are 
the generators of the 
symmetric algebra of the dual of $\bigoplus_n u(1)\subset su(n+1)$.

Note that the seemingly infinite product
should read an abbreviated form of the finite product:
for two partitions $\R_i,\R_j$, 
\begin{equation}\label{ab}
\begin{split}
&\prod_{k,l=1}^{\infty}
\frac{a_{i,j}+\hbar(\mu_{i,k}-\mu_{j,l}+l-k)}{a_{i,j}+\hbar(l-k)}
\\
&=
\prod_{(k,l)\in\R_i}\frac{1}
{a_{i,j}+\hbar(\mu_{i,k}-l+\mu_{j,l}^{\vee}-k+1)}
\prod_{(k,l)\in\R_j}\frac{1}
{a_{i,j}+\hbar({-\mu_{i,l}^{\vee}+k-\mu_{j,k}+l-1})}.
\end{split}
\end{equation}
Here $\mu_{i,k}^{\vee}$ (resp. 
$\mu_{j,k}^{\vee}$) is the $k$-th part of 
$\conj{\R_i}$ (resp. $\conj{\R_j}$), i.e.
$\conj{R_i}=(\mu_{i,1}^{\vee},\mu_{i,2}^{\vee},\ldots)$.
$(k,l)\in\R$ means that there is a box in the Young diagram $\R$
at the place of $k$-th row and $l$-th column.
It becomes important later
that all the factors appear only in the denominator,
not in the numerator
when we derive the 
asymptotic form of the Gopakumar--Vafa invariants.

Nekrasov's partition function is 
invariant under the action of the Weyl group of $A_n$ 
(the symmetric group $S_{n+1}$). It is also invariant
under the ${\bf Z}_2$ action, which is generated by
$(a_{1,2},\ldots,a_{n,n+1})\to(-a_{1,2},\ldots,-a_{n,n+1})$
(this ${\bf Z}_2$ action coincides with the Weyl group action
in the case of  $A_1$).
Some of these symmetries will later appear in the
result of the 
asymptotic form of the Gopakumar--Vafa invariants.

\section{\bf $\boldsymbol{A_n}$-fibration over ${\bf P}^1$}
\label{toric}
\hspace*{2.5ex}
In this section we describe the
$A_n$-fibration over ${\bf P}^1$ ($n\geq 1$).
By this term, we  mean  
the smooth, Calabi--Yau
(i.e., the canonical bundle of which is trivial),
toric variety of complex three dimensions
which has the structure of the fibration of 
the minimal resolution of the 
$A_n$-singularity over ${\bf P}^1$.

There exist $(n+2)$ different 
such Calabi--Yau toric threefolds.
We label them by an integer $m$ $(-n+1\leq m\leq 2)$
and call it $X_{A_n}^m$. 
$X_{A_n}^m$ is specified by 
the polytope 
\begin{equation}\label{polytope}
\triangle_{A_n}^{m}=[(0,1),(-m,-1),(-1,0),(n,0)]\subset {\bf R}^2 
\end{equation}
with the following triangulation;
let $v_i$ $(1\leq i\leq n+4)$ be
the integral points:
\begin{equation}
v_1=(0,1),v_2=(-m,-1),v_3=(-1,0),v_4=(0,0),\ldots,v_{n+4}=(n,0);
\end{equation}
then the triangulation is such that its 2-simplices are
$[v_1,v_i,v_{i+1}]$  and
$[v_2,v_i,v_{i+1}]$ $(3\leq i\leq n+3)$.
(see Figure \ref{a2graph} for $n=2$ example).
Here 
$[v_j,\ldots,v_k]$
is
the convex hull of the vectors $v_j,\ldots,v_k$.
The fan for $X_{A_n}^m$ is such that 
its section at the height 1 is the polytope 
$\triangle_{A_n}^m$.

A basis of $H_2(X_{A_n}^m;{\bf Z})$ consists of
the homology classes of
the base space ${\bf P}^1=:C_B$, and $n$-exceptional 
curves of the fiber space  
$C_j(1\leq j\leq n)$.
We can calculate 
$H_2(X_{A_n}^m;{\bf Z})$ 
using the spectral sequence argument \cite{Fulton} (chapter 2)
and the result is as follows.
Each interior point corresponds to a compact surface
and such surfaces generate $H_4(X_{A_n}^m;{\bf Z})$ freely.
Each interior edge in $\triangle_{A_n}^m$ 
corresponds to a torus
invariant curve ${\bf P}^1$
and such ${\bf P}^1$'s generate $H_2(X_{A_n}^m;{\bf Z})$.
Let $C_{ij}$ be the ${\bf P}^1$ corresponding to the
interior edge spanned by $v_i$ and $v_j$.
Then there are two relations for each interior point $v_i$:
\begin{equation}
\sum_j(v_j-v_i)[C_{ij}]=\vec{0} \qquad 
(4\leq i\leq n+3).
\end{equation}
Here the summation is over $j$ such that 
$v_i$ and $v_j$ span a 1-simplex.
Therefore
$[C_B]$ corresponds to the edge
$[v_3,v_4]$ for $m=1,2$,
to both $[v_{r+3},v_{r+4}]$ and $[v_{r+4},v_{r+5}]$ 
for $m=-2r \big(0\leq r\leq [\frac{m-1}{2}]\big)$,
to $[v_{r+3},v_{r+4}]$ for 
$m=-2r+1\big(1\leq i\leq [\frac{m}{2}]\big)$;
the exceptional curve
$[C_i]$ corresponds to
$[v_1,v_{i+3}]$ and $[v_2,v_{i+3}]$ $(1\leq i\leq n)$
(more precisely, we should say that we define the order $i$
of $C_i$ as such).

The
generating function of the Gromov--Witten invariants for 
$A_n$-fibration over ${\bf P}^1$ denoted by $X_{A_n}^m$, takes 
the following form:
\begin{equation}
\begin{split}\label{gwform}
{\mathcal F}_{\rm GW}(X_{A_n}^m)
&=\sum_{g=0}^{\infty}{g_s}^{2g-2}
{{\mathcal F}^g_{\rm GW}}(X_{A_n}^m),
\\
{\mathcal F}^g_{\rm GW}(X_{A_n}^m)
&=
\sum_{d_{\rm B},d_1,\ldots,d_n\geq 0}
N_{g,d_{\rm B},d_1,\ldots,d_n}
{q_{\rm B}}^{d_{\rm B}}{q_1}^{d_1}\cdots{q_n}^{d_n}.
\\
q_{\rm B}&=e^{-t_{\rm B}},
\quad q_i=e^{-t_i} (1\leq i\leq n).
\end{split}
\end{equation}
$N_{g,d_{\rm B},\ldots,d_n}$ is the genus $g$, 0-pointed
Gromov--Witten invariant of $X_{A_n}^m$
for the homology class $d_{\rm B}[C_B]+d_1[C_1]+\cdots+d_n[C_n]$.
$t_B$ (resp. $t_i$) is the 
K\"ahler parameter for the 2-cycle
$[C_B]$ (resp. $[C_i]$ $(1\leq i\leq n)$).
$g_s$ is the  variable for the genus expansion,
and  is identified with the string coupling.

Gromov--Witten invariants are generically rational numbers,
but the generating function can be expressed in terms of
integral invariants called
the Gopakumar--Vafa invariants
\cite{GoVa2, GoVa3}.
In the case of the $A_n$-fibration over ${\bf P}^1$,
the generating function is written
as follows:
\begin{equation}\label{gvform}
\begin{split}
{\mathcal F}_{\rm GW}(X_{A_n}^m)
&=\sum_{g=0}^{\infty}
\sum_{d_{\rm B},d_1,\ldots,d_n\geq 0}
\sum_{k=1}^{\infty}
\frac{n_{d_{\rm B},d_1,\ldots,d_n}^g}{k}
\Big(2\sin\frac{k g_s}{2}\Big)^{2g-2}
\big({q_{\rm B}}^{d_{\rm B}}{q_1}^{d_1}\cdots{q_n}^{d_n}\big)^k.
\end{split}
\end{equation}
Here $n_{d_{\rm B},d_1,\ldots,d_n}^g$
is the Gopakumar--Vafa invariant of $X_{A_n}^m$
for the homology class $d_{\rm B}[C_B]+d_1[C_1]+\cdots+d_n[C_n]$
and for genus $g$.

\section{Topological String Amplitudes}
\label{iqka}
\hspace*{2.5ex}
In this section 
we briefly review some of
Iqbal and Kashani-Poor's results:
the topological string amplitude
for $A_n$-fibration over ${\bf P}^1$
and its field theory limit \cite{IqKa1}.


It is conjectured that the topological string amplitude
obtained by the geometric transition and the Chern--Simons theory
gives the generating function of Gromov--Witten invariants
\cite{Marino}.
For the $A_n$-fibration over ${\bf P}^1$, 
\begin{equation}\label{pf_gf}
\log{\mathcal Z}(X_{A_n}^m)|_{q=e^{ig_s}}
={\mathcal F}_{\rm GW}(X_{A_n}^m).
\end{equation}
Here $q$ is a parameter which should be
identified with $\exp \big(\frac{2\pi i}{N+k}\big)$ in  
the $SU(N)$ Chern--Simons theory but in this 
context it is just a formal variable.
$i$ denotes the imaginary unit.
$g_s$ is the parameter of the genus expansion
appeared in (\ref{gwform})(\ref{gvform}).

Iqbal and Kashani-Poor derived the
topological string amplitude 
using certain identities on the 
summation over partitions
\cite{IqKa1,IqKa2}. 
The proof of the identities appeared later in the paper
by Zhou \cite{Zhou} (Theorem 8.1).
The topological string amplitude is 
\begin{align}
\nonumber
{\mathcal Z}(X_{A_n}^m)&=
{Z_{d_{\rm B}=0}}\;{Z_{d_{\rm B}\geq 1}},\\
\label{zdb0}
Z_{d_{\rm B}=0}&:=\prod_{1\leq i<j\leq n+1}K(q_{i,j})^2,\\
\label{zdbgeq}
Z_{d_{\rm B}\geq 1}&:=
\sum_{\R_1,\ldots,\R_{n+1}}\Big[
(-1)^{m\sum_{i=1}^{n+1}\w{\R_i}}\,
q^{\sum_{i=1}^{n+1}\frac{m-4+2i}{2}\kappa(\R_i)}
{q_{\rm B}}^{\sum_{i=1}^{n+1}\w{\R_i}}
\\\nonumber&\times
\prod_{1\leq i\leq [\frac{-m+1}{2}]}
{q_i}^{(-m+2-2i)\sum_{k=1}^i\w{\R_k}}
\prod_{[\frac{-m+1}{2}]<i\leq n}
\,
{q_i}^{(m-2+2i)\sum_{k=i+1}^{n+1}\w{\R_k}}
\\\nonumber&\times
\prod_{1\leq i\leq n+1}W_{\R_i}(q)^2
\prod_{1\leq i<j\leq n+1}g_{\conj{\R_i},\R_j}(q_{i,j},q)^2
\Big].
\end{align}
Here $\R_1,\ldots,\R_{n+1}$ are partitions,
$q_{i,j}:=\prod_{k=i}^{j-1}q_k$.
$\kappa(\R)=\sum_{i=1}^{\len{\R}}\mu_i(\mu_i-2i+1)$.
\begin{align}\label{kx}
K(x)&:=\exp\Big[\sum_{k=1}^{\infty}
\frac{q^k}{k(q^k-1)^2}x^k
\Big].\\
\label{wr}
W_{\R}(q)&:=q^{\frac{\kappa(\R)}{4}}
\prod_{1\leq i<j\leq \len{\R}}
\frac{[\mu_i-\mu_j+j-i]}{[j-i]}
\prod_{i=1}^{\len{\R}}
\prod_{j=1}^{\mu_i}\frac{1}{[j-i+\len{\R}]},
\quad \big([x]:=q^{\frac{x}{2}}-q^{-\frac{x}{2}}\big),\\
\label{gr}
g_{\R_1,\conj{\R_2}}(x;q)&=
\prod_{(i,j)\in\R_1}\frac{1}
{(1-x q^{\mu_{1,i}-j+\mu_{2,j}^{\vee}-i+1})}
\prod_{(i,j)\in\R_2}\frac{1}
{(1-x q^{-\mu_{1,j}^{\vee}+i-\mu_{2,i}+j-1})}.
\end{align}
Note that we can deal with 
the Gopakumar--Vafa invariants with 
$d_{\rm B}=0$ and $d_{\rm B}\geq 1$ separately.
$Z_{d_{\rm B}=0}$ gives 
the Gopakumar--Vafa invariants with $d_{\rm B}=0$ 
while
${Z_{d_{\rm B}\geq1}}$ gives those 
with  $d_{\rm B}\geq 1$.
This is because $Z_{d_{\rm B}=0}$ does not depend on $q_{\rm B}$
and $Z_{d_{\rm B}\geq 1}=1+{\mathcal O}(q_B)$. 
From $Z_{d_{\rm B}=0}$ one can easily read the
Gopakumar--Vafa invariants for $d_{\rm B}=0$:
$n^g_{d_{\rm B}=0,d_1,\ldots,d_n}=-2$ when
$d_i=d_{i+1}=\cdots=d_j=1$ $(1\leq i\leq j\leq n)$
and other $d_k$'s are zero,
and  $n^g_{d_{\rm B}=0,d_1,\ldots,d_n}=0$ otherwise.
In other words, the  value of 
$n^{g=0}_{d_{\rm B}=0,d_1,\ldots,d_n}$  is nonzero only
for a homology class corresponding to a positive root
under the identification of 
$[C_1],\ldots,[C_n]$ with the simple roots of
the Lie algebra $A_n$.

$Z_{d_{\rm B}\geq 1}$ in (\ref{zdbgeq}) 
is also written in the following form:
\begin{equation}
\begin{split}\label{IK_pf2}
Z_{d_{\rm B}\geq 1}&=
\sum_{\R_1,\ldots,\R_{n+1}}
\Big[
(-1)^{(n+m+1)\sum_{i=1}^{n+1}\w{\R_i})}
2^{-2(n+1)\sum_{i=1}^{n+1}\w{\R_i}}
q^{\frac{n+m-1}{2}
\sum_{i=1}^{n+1}\kappa(\R_i)}
\\&\times
{q_{\rm B}}^{\sum_{i=1}^{n+1}\w{\R_i}}
\prod_{1\leq k\leq [\frac{-m+2}{2}]}
{q_k}^{(-m-n+1-k)(\w{\R_1}+\cdots+\w{\R_k})-
k(\w{\R_{k+1}}+\cdots+\w{\R_{n+1}})
}
\\&\times
\prod_{ [\frac{-m+2}{2}]<k\leq n}
{q_k}^{(-n-1+k)(\w{\R_1}+\cdots+\w{\R_k})+
(m-2+k)(\w{\R_{k+1}}+\cdots+\w{\R_{n+1}})
}
\\
&\times
\prod_{1\leq i,j\leq n+1}\prod_{k,l=1}^{\infty}
\frac{\sinh\frac{\beta}{2}(a_{i,j}+\hbar(\mu_{i,k}-\mu_{j,l}+l-k))}
{\sinh \frac{\beta}{2}(a_{i,j}+\hbar(l-k))}\Big].
\end{split}
\end{equation}
Here $\beta,a_i,\hbar$ are introduced as
\begin{equation}
q_i=e^{-\beta(a_i-a_{i+1})}
,\qquad
q=e^{-\beta{\hbar}},
\end{equation}
and $a_{i,j}:=a_i-a_j$.
The last product in (\ref{IK_pf2}) 
is equal to the factor that appears in the 
Nekrasov's complete string partition function
\cite{Nek, IqKa1}.
The product is the finite product in the same manner as
we mentioned in section \ref{nekrasov}.

The {\it field theory limit} is the limit $\beta\to 0$ with
\begin{equation}\label{limit}
q_{\rm B}=(-1)^{n+m+1}(\beta\Lambda)^{2(n+1)},\quad
t_i={-\beta a_{i,i+1}} (1\leq i\leq n),\quad
q=e^{-\beta{\hbar}}.
\end{equation}
Taking the field theory limit, $Z_{d_{\rm B}\geq 1}$ becomes
\begin{equation}\label{limit2}
\lim_{\beta\to0} 
Z_{d_{\rm B}\geq 1}= {\mathcal Z}_{\rm Nekrasov}^{A_n}.
\end{equation}
Therefore, the field theory limit is the limit where
the amplitude of the four-dimensional theory
is reproduced from  the topological strings.

Let us summarize the correspondence
between the parameters in 
the topological string amplitude
and those in the four-dimensional
gauge theory (Nekrasov's partition function).
The K\"ahler parameters $t_i(1\leq i\leq n)$ of 
the fiber are proportional to the vacuum expectation 
value of the complex scalar $a_{i,i+1}$ 
in the gauge theory in four dimensions.
The genus expansion parameter $g_s$ 
is proportional to the  parameter $\hbar$.
In the recent work of Eguchi and Kanno \cite{EgKa},
the parameter $\beta$ is identified with the 
radius of the fifth-dimensional circle  in the 
five-dimensional gauge theory.

The list below is the identification of the notation 
of Iqbal--Kashani-Poor \cite{IqKa1} with ours.
\begin{equation*}
\begin{array}{rcl}\hline
\mbox{Iqbal--Kashani-Poor \cite{IqKa1}}&&\mbox{ Here }\\\hline
\prod_{k}(1-q^k x)^{-C_{k}(\R_1,\R_2)}&&g_{\R_1,\R_2}(x,q)\\
N&&n+1\\
m&&m+n-1\\
t_{F_i}(Q_{F_i})&&t_{n+1-i}(q_{n+1-i})
\\
\{a_1,\ldots,a_{N-1}\}&&\{a_n,\ldots,a_1\}\\
\{R_{1},\ldots,R_N\}&&\{\R_{n+1},\ldots,\R_{1}\}
\\\hline
\end{array}
\end{equation*}
\remark : The relation between the three point vertex amplitudes
in \cite{IqKa1} and those appeared in the
recent paper of Aganagic et al.
\cite{AgKlMaVa} is $V_{\R_1,\R_2,\R_3}=
q^{\frac{\kappa(\R_3)}{2}}C_{\R_1,\R_2,\conj{\R_3}}$.

\section{Asymptotic Form of the Gopakumar--Vafa Invariants}
\label{gv}
\hspace*{2.5ex}
In this section,
we derive the asymptotic form $r_{d_{\rm B}}^{(g)}(d_1,\ldots,d_n)$
of the Gopakumar--Vafa
invariants $n_{d_{\rm B},d_1,\ldots,d_n}^g$ for $d_B\geq 1$.

Let us state the result first: the asymptotic form is
given by
\begin{equation}
\label{asymptotics}
r_{d_{\rm B}}^{(g)}(d_1,\dots,d_n)=
(-1)^{(n+m+1)d_{\rm B}}{{\mathcal L}_{d_1,a_{1,2}}}^{-1}\circ
\cdots\circ{{\mathcal L}_{d_n,a_{n,n+1}}}^{-1}
{{\mathcal F}_{d_{\rm B}}^{(g)}}(a_{1,2},\ldots,a_{n,n+1}).
\end{equation}
This formula holds 
in the region
\begin{equation}\label{region}
d_{\rm B}\geq 1, \quad
d_1,\ldots, d_n\gg d_{\rm B}(g+1).
\end{equation}
Here
${\mathcal L}_{(x,s)}$ is the Laplace transform, i.e. 
${\mathcal L}_{(x,s)}\big(f(x)\big)=\int_{0}^{\infty}\,dx\; e^{-sx}f(x)$
and ${{\mathcal L}_{(x,s)}}^{-1}$ is its inverse.
${\mathcal F}_{k}^{(g)}(a_{1,2},\ldots,a_{n,n+1})$
is defined from the 
Nekrasov's partition function as follows:
\begin{equation} 
\label{rhs}
\log {{\mathcal Z}_{\rm Nekrasov}^{A_n}}
=\sum_{g=0}^{\infty}(i\hbar)^{2g-2}
\sum_{k=1}^{\infty}{\Lambda^{2(n+1)k}}
{{\mathcal F}_{k}^{(g)}(a_{1,2},\ldots,a_{n,n+1})}.
\end{equation}
It should be regarded as the function in $n$ variables
$a_{1,2},\ldots,a_{n,n+1}$.

There are several remarks.
\\
1. ${\mathcal F}_{k}^{(g)}(a_{1,2},\ldots,a_{n,n+1})$ 
takes the following form
\begin{equation}
\label{rhs2}
{\mathcal F}_{k}^{(g)}(a_{1,2},\ldots,a_{n,n+1})
=\sum_{\sum k_{ij}=2(n+1)k+2g-2}
A_{\{k_{ij}\}}\prod_{1\leq i\leq j\leq n}
\frac{1}{{a_{i,j}}^{k_{ij}}}
\end{equation}
where $A_{\{k_{ij}\}}$ is a constant prefactor.
The counting of the 
degree $-2(n+1)k-2g+2$ in ${a_{i,j}}$'s in the right-hand side 
is as follows.
Assign the degree 1 to both
$a_{i,j}$'s and $\hbar$
in Nekrasov's partition function (\ref{nek}).
Then the term
with $\w{\R_1}+\cdots+\w{\R_{n+1}}=k$
should have
the degree $-2(n+1)k$. 
Subtracting the degree $ 2g-2$ of $\hbar$, 
we obtain the degree $-2(n+1)k-2g+2$.
Note that 
we can perform the inverse Laplace transform because
$a_{i,j}$'s do not appear in the numerator in (\ref{rhs2}).
\\
2. The appearance of the inverse Laplace transform 
 is a very natural consequence.
If we just rewrite $q_i$ as $e^{-t_i}$ $(1\leq i\leq n)$ in 
the generating function of the Gromov--Witten invariants
(\ref{gwform}), it becomes
\begin{equation}
\begin{split}\label{part1}
\sum_{d_{\rm B}\geq 1} {q_{\rm B}}^{d_{\rm B}}
\sum_{g=0}^{\infty}{g_s}^{2g-2}
\sum_{d_1,\ldots,d_n\geq 0}
N_{g,d_{\rm B},d_1,\ldots,d_n}
{e^{-t_1 d_1}}\cdots{e^{-t_n d_n}}.
\end{split}
\end{equation}
We have shown only the relevant part for  
$N_{g, d_B,d_1,\ldots,d_n}$ with $d_B\geq 1$ here.
Then if we assume that 
the summation over $d_1,\cdots,d_n$ 
can be replaced with the integration,
we can regard the integral as the 
Laplace transform of $N_{g,d_{\rm B},d_1,\ldots,d_n}$
from the variables $d_1,\cdots,d_n$ 
to the variables $t_1,\cdots,t_n$.
Given that $t_i$ is proportional to $a_{i,i+1}$ (\ref{limit})
and that  the inverse Laplace 
transform exists as  mentioned above, 
the asymptotic form of
the Gromov--Witten invariants is given by 
the inverse Laplace transform.
\\
3. $r_{d_{\rm B}}^{(g)}(d_1,\ldots,d_n)$ 
is the homogeneous
polynomial in $d_1,\ldots,d_n$
of degree 
\begin{equation}\label{degree}
2(n+1)d_{\rm B}+2g-2-n.
\end{equation}
Note that 
the degree grows as $d_{\rm B}$ and $g$ grow.
($\because$)
Recall that 
performing the inverse Laplace transform once
reduces the degree by 1 (for example,
${\mathcal L}_{x,s}^{-1}\big(\frac{1}{s^n}\big)$
is a polynomial of degree $n-1$ in $x$). 
Therefore the degree of the result of the
inverse Laplace transform 
for $A_n$ is smaller by $n$ than 
the total degree $2(n+1)k+2g-2$
of ${a_{i,j}}^{-1}$'s in 
${\mathcal F}_{k}^{(g)}(a_{1,2},\ldots,a_{n,n+1})$
(\ref{rhs2}). 
\\
4. The condition $d_1,\ldots,d_n\gg 1$ is necessary
so that we can replace the summation with 
the integration. 
The condition $d_1,\ldots,d_n\gg d_{\rm B}(g+1)$ (\ref{region})
is further
required so that we can  neglect the contribution from
multiple covering and the bubbling.
\\
5. $r_{d_{\rm B}}^{(g)}(d_1,\ldots,d_n)$ 
is invariant under $(d_1,d_2,\ldots,d_n)\mapsto
(d_n,d_{n-1},\ldots,d_1)$.
It is because $Z_{d_{\rm B}\geq1}$ (\ref{zdbgeq})
is invariant under
\begin{equation}
(q_1,q_2,\ldots,q_n)\mapsto
(q_n,q_{n-1},\ldots,q_1).
\end{equation} 
{It is possible that
the  Weyl invariance manifests itself
in the asymptotic form in other manners.
For $A_2$ cases, see the remark in section \ref{exa2}.}

\vsp

\noindent (Proof) 
\\
We  consider the 
asymptotic form of the Gromov--Witten invariants 
$N_{g,d_{\rm B},d_1,\ldots,d_n}$ first
and that of the Gopakumar--Vafa invariants 
$n_{d_{\rm B},d_1,\ldots,d_n}^g$ next.
As it turned out, 
the two are the same,
because we can neglect the bubbling effect and 
the multicovering effect when $d_1,\ldots,d_n\gg 
d_{\rm B}(g+1)$.

We use the notations
$\va:=(a_{1,2},\ldots,a_{n,n+1})$,
$\vd:=(d_1,\ldots,d_n)$,
$\vt:=(t_1,\ldots,t_n)$
in the rest of this section.

Recall that 
the part  $Z_{d_{\rm B}\geq1}$
becomes equal to Nekrasov's partition function 
in the field theory limit  (\ref{limit2}).
Let us consider the logarithm of the equation.
Then its right-hand side  is  written as
(\ref{rhs}).
On the other hand, by
substituting (\ref{limit}) into (\ref{part1}),
the left-hand side  is written as follows:
\begin{equation}\label{lhs}
\begin{split}
\lim_{\beta\to 0}
\log Z_{d_{\rm B}\geq 1}&=
\sum_{g=0}^{\infty}
\big(i\hbar\beta\big)^{2g-2}
\sum_{d_{\rm B}=1}^{\infty}
\big((-1)^{n+m+1}(\beta\Lambda)^{2(n+1)}\big)^{d_{\rm B}}
\\
&\times\sum_{\vd}
N_{g,d_{\rm B},\vd} e^{-\vd\cdot\vt}
|_{\vt=\beta\va}.
\end{split}
\end{equation}
Comparing the right-hand side (\ref{rhs}) and 
the left-hand side (\ref{lhs}) as the
formal power series in $\hbar$ and $\Lambda$,
the following holds up to the lowest order in $\beta$:
\begin{equation}\label{sum}
{\mathcal F}_k^{(g)}(\va)=(-1)^{(n+m+1)k}\beta^{2(n+1)k+2g-2}
\sum_{\vd}N_{g,d_{\rm B}=k,\vd}
e^{-\vd\cdot\vt}
|_{\vt=\beta\va}.
\end{equation}
Now we  replace the sum over $d_1,\ldots,d_n$ with
the integration in the right-hand side. 
Then, the integral is nothing but the Laplace transform of 
$N_{g,d_{\rm B},\vd}$ from the variables $(d_1,\ldots,d_n)$ 
to $(t_1,\ldots,t_n)$:
\begin{equation}
{\mathcal F}_{d_{\rm B}}^{(g)}(\va) 
\sim (-1)^{(n+m+1)d_{\rm B}}\beta^{2(n+1)d_{\rm B}+2g-2}
{\mathcal L}_{(d_1,t_1)}\circ\cdots\circ
{\mathcal L}_{(d_n,t_n)}
(N_{g,d_{\rm B},\vd}).
\end{equation}
Therefore the asymptotic form of the 
Gromov--Witten invariant $N_{g,d_{\rm B},\vd}$ 
for given 
$g$ and $d_{\rm B}$ is obtained by the inverse Laplace transform
of ${\mathcal F}_{d_{\rm B}}^{(g)}(\va)$:
\begin{equation}\label{bef}
N_{g,d_{\rm B},\vd}\sim
(-1)^{(n+m+1)d_{\rm B}}\beta^{2(n+1)d_{\rm B}+2g-2}
{\mathcal L}_{(d_1,t_1)}^{-1}\circ\cdots\circ
{\mathcal L}_{(d_n,t_n)}^{-1}
{\mathcal F}_{d_{\rm B}}^{(g)}(\va)|_{\va=\vt/\beta}
\end{equation}
Since ${\mathcal F}_{d_{\rm B}}^{(g)}(\va)$
is homogeneous in $1/a_{i,j}$'s with 
degree  $2(n+1)d_{\rm B}+2g-2$,
${\mathcal F}_{d_{\rm B}}^{(g)}(\va)
|_{\va=\vt/\beta}=
\beta^{-2(n+1)d_{\rm B}-2g+2}{\mathcal F}_{d_{\rm B}}^{(g)}(\vt)
$.
Hence the powers of $\beta$ in (\ref{bef})
vanish and the result
is finite at $\beta\to 0$.
Thus, just rewriting $\vt$ with $\va$, we obtain
\begin{equation}
\begin{split}
N_{g,d_{\rm B},\vd}&\sim
(-1)^{(n+m+1)d_{\rm B}}
{\mathcal L}_{(a_{1,2},t_1)}^{-1}\circ\cdots\circ
{\mathcal L}_{(a_{n,n+1},t_n)}^{-1}
{\mathcal F}_{d_{\rm B}}^{(g)}(\va)
=:r_{d_{\rm B}}^{(g)}(\vd).
\end{split}
\end{equation}
When we have replaced the summation with the integration
in (\ref{sum}), we have assumed that 
the contribution from the 
Gromov--Witten invariants with large 
values of $d_1,\cdots,d_n$ is dominant.
The assumption could be justified by this result
($\because$ (\ref{degree})).

Next we consider the asymptotic form of the 
Gopakumar--Vafa invariants $n_{d_{\rm B},\vd}^g$.
The Gopakumar--Vafa invariants are 
written in terms  the Gromov--Witten invariants
as follows \cite{BrPa}:
\begin{equation}\label{gw_gv}
n_{d_{\rm B},\vd}^g=
\sum_{g'=0}^g\sum_{k|(d_{\rm B},\vd)}
\mu(k)k^{2g'-3}\alpha_{g,g'}N_{g',d_{\rm B}/k,\vd/k}
\end{equation}
where $\alpha_{g,g'}$ is the coefficient of $r^{g-g'}$
in the series
\begin{equation*}
\Big(\frac{\arcsin(\sqrt{r}/2)}{\sqrt{r}/2}\Big)^{2g'-2}.
\end{equation*}
Note that
the degree  of $r_{d_{\rm B}}^{(g)}(\vd)$ is 
$2(n+1)d_{\rm B}+2g-2-n$ (\ref{degree}) and 
grows with
$g$ and $d_{\rm B}$. 
Therefore  $N_{g',d_{\rm B}/k,\vd/k}$ with 
$k\geq 2$ or $g'<g$ is sufficiently smaller
when $d_1,\ldots,d_n\gg d_{\rm B}$:
\begin{equation*}
N_{g',d_{\rm B}/k,\vd/k}\ll N_{g,d_{\rm B},\vd}
\quad\mbox{ if } \quad
g'<g \mbox{ or } k>1.
\end{equation*}
And the number of such terms in the right-hand side of
(\ref{gw_gv}) is 
at most $d_{\rm B}(g+1)$.
Thus, the contribution to the Gopakumar--Vafa invariant
from the Gromov--Witten invariants
with lower genera  (bubbling effect)
and lower degrees (multiple covering)
can be neglected
if $d_1,\ldots,d_n\gg d_{\rm B}(g+1)$.
When this condition is satisfied,
\begin{equation}
N_{g,d_{\rm B},\vd}
\sim n_{d_{\rm B},\vd}^g
\sim r_{d_{\rm B}}^{(g)}(\vd).
\end{equation}

\section{Example: $\boldsymbol{A_2}$-fibration 
over ${\bf P}^1$}\label{exa2}
\begin{figure}[t]
\begin{center}
\scriptsize{
\input{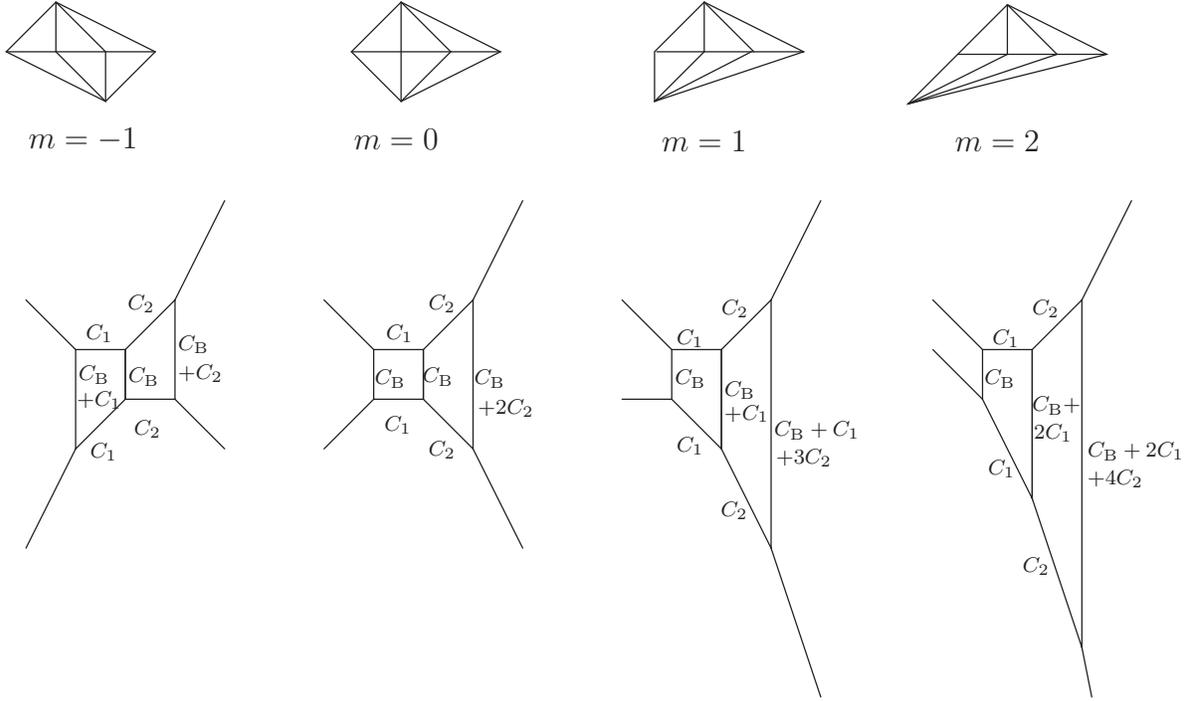}
}
\end{center}
\caption{The
section $\triangle_{A_2}^m$ of the fan at the height 1 (top)
and the web diagram $\Gamma$ 
(bottom)
for $A_2$-fibration over ${\bf P}^1$. 
}
\label{a2graph}
\end{figure}

\begin{figure}[t]
\includegraphics[width=8cm]{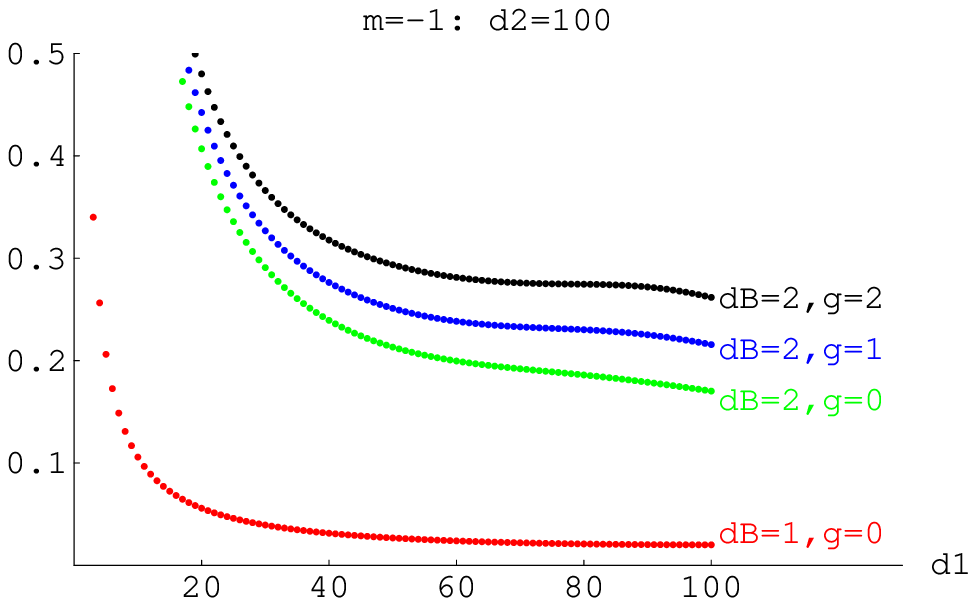}
\includegraphics[width=8cm]{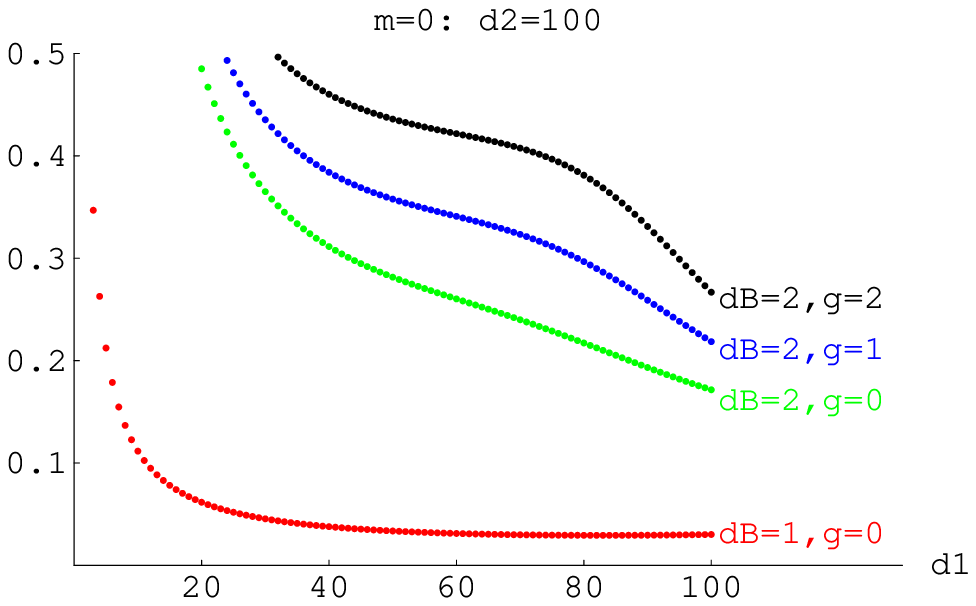}

\vspace*{1cm}
\includegraphics[width=8cm]{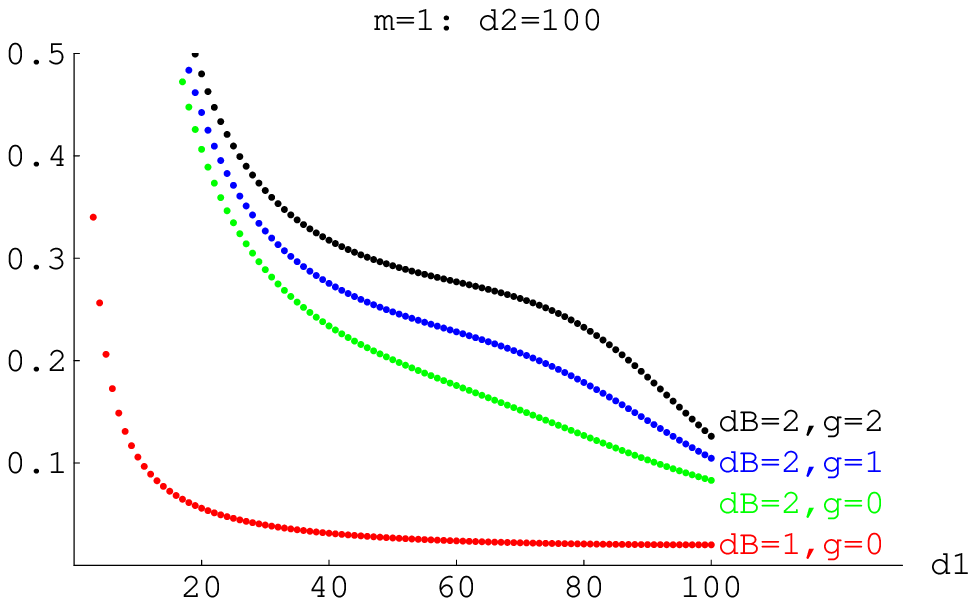}
\includegraphics[width=8cm]{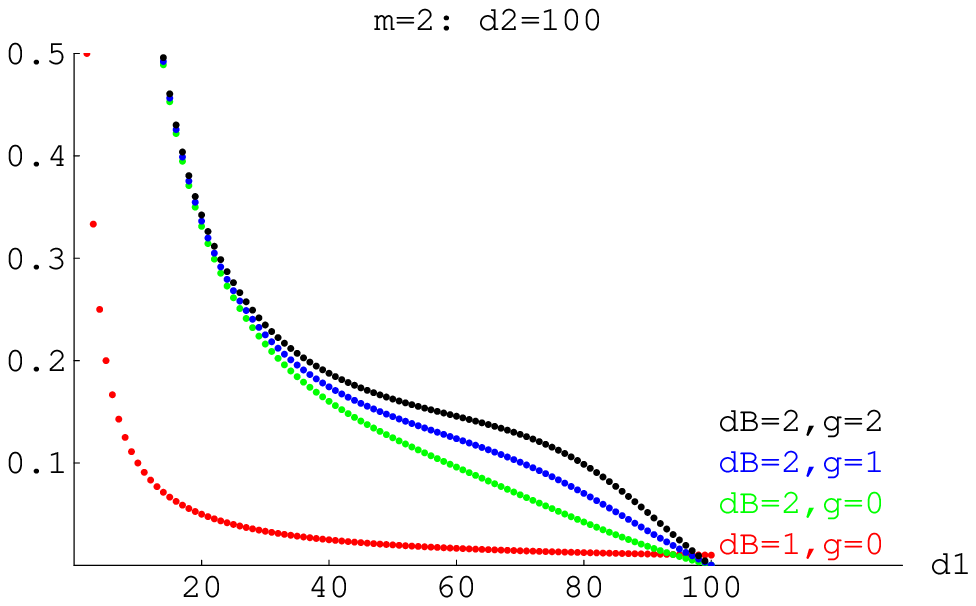}
\caption{The 
ratio between the Gopakumar--Vafa invariants 
$n_{d_B,d_1,d_2}^g$
and the asymptotic form $r^{(g)}_{d_B}(d_1,d_2)$.
The vertical axis of the plots is
$n_{d_B,d_1,d_2}^g/r^{(g)}_{d_B}(d_1,d_2)-1$.
}\label{plot1}
\end{figure}

\hspace*{2.5ex}
The asymptotic form for the  $A_1$ case at genus zero
was studied in \cite{KaKlVa, KoNa}.
In this section we  study the case $n=2$.
There are four Calabi--Yau toric threefolds
which has the structure of the fibration of   
$A_2$-singularity over ${\bf P}^1$
(figure \ref{a2graph}).
We  look into $m=-1$ and $g=0,d_{\rm B}=1,2$ in detail
for the illustrative purpose. 
For more thorough results, see appendix \ref{data}.
First we calculate 
the asymptotic form $r_{d_{\rm B}}^{(g)}(d_1,d_2)$ (\ref{asymptotics})
from Nekrasov's partition function.
Then we calculate the Gopakumar--Vafa 
invariants from the topological string amplitude of 
Iqbal--Kashani-Poor (\ref{zdbgeq}) \cite{IqKa1}.
(We have checked that the Gopakumar--Vafa invariants
from (\ref{zdbgeq}) agree with those obtained from 
the local B-model calculation
for all four $A_2$-fibration over ${\bf P}^1$,
up to $d_{\rm B}\leq 2$, $d_1,d_2\leq 21$.
The data of the local B-model is presented in the appendix
\ref{localB}.)
Finally we  see that the ratio
approaches 
to one when  $d_1,d_2$ are large (figure \ref{plot1}).

\vsp
\noindent{\bf Asymptotic form}

In section \ref{gv},
we have derived the asymptotic form $r_{d_{\rm B}}^{(g)}(d_1,d_2)$
of the Gopakumar--Vafa invariants $n_{d_{\rm B},d_1,d_2}^g$.

\begin{equation}\label{a2specialized}
\begin{split}
&{\mathcal Z}_{{\rm Nekrasov}}^{A_2}=
\sum_{\R_1,\R_2,\R_{3}}
\Lambda^{\w{\R_1}+\w{\R_2}+\w{\R_{3}}}
\prod_{i,j=1}^{3}\prod_{k,l=1}^{\infty}
\frac{a_{i,j}+\hbar(\mu_{i,k}-\mu_{j,l}+l-k)}{a_{i,j}+\hbar(l-k)}.
\\
&\log{\mathcal Z}_{{\rm Nekrasov}}^{A_2}=
\sum_{g=0}^{\infty}(\sqrt{-1}h)^{2g-2}\sum_{k=1}^{\infty}
\mathcal{F}^{(g)}_k(a_{1,2},a_{2,3})\Lambda^{6k}.
\\
&r_{d_{\rm B}}^{(g)}(d_1,d_2)=(-1)^{(n+m+1)d_{\rm B}}
{\mathcal L}^{-1}_{d_1,a_{1,2}}\circ
{\mathcal L}^{-1}_{d_2,a_{2,3}}
{\mathcal F}_{d_{\rm B}}^{(g)}(a_{1,2},a_{2,3}).
\end{split}
\end{equation}
For instance, an explicit expression for
first few ${\mathcal F}_{k}^{(g)}$'s are as follows:
\begin{equation}\label{swa2}
\begin{split}
{\mathcal F}_{1}^{(0)}&
={\frac{1}{{{{a_{1,2}}}^2}{{{a_{1,3}}}^2}}}+
{\frac{1}{{{{a_{1,2}}}^2}{{{a_{2,3}}}^2}}}+ 
{\frac{1}{{{{a_{1,3}}}^2}{{{a_{2,3}}}^2}}},
\\
{\mathcal F}_{2}^{(0)}&
={\frac{3}{2{{{a_{1,2}}}^4}{{{a_{1,3}}}^6}}}+ 
{\frac{3}{2{{{a_{1,2}}}^6}{{{a_{1,3}}}^4}}}+ 
{\frac{3}{2{{{a_{1,2}}}^4}{{{a_{2,3}}}^6}}}+ 
{\frac{3}{2{{{a_{1,3}}}^4}{{{a_{2,3}}}^6}}}+
{\frac{3}{2{{{a_{1,2}}}^6}{{{a_{2,3}}}^4}}}+ 
{\frac{3}{2{{{a_{1,3}}}^6}{{{a_{2,3}}}^4}}} 
\\&+{\frac{2}{{{{a_{1,2}}}^5}{{{a_{1,3}}}^5}}}
- {\frac{2}{{{{a_{1,2}}}^5}{{{a_{2,3}}}^5}}} + 
{\frac{2}{{{{a_{1,3}}}^5}{{{a_{2,3}}}^5}}} 
\\&+ 
{\frac{2}{{{{a_{1,2}}}^2}{{{a_{1,3}}}^2}{{{a_{2,3}}}^6}}}
+{\frac{2}{{{{a_{1,2}}}^2}{{{a_{1,3}}}^6}{{{a_{2,3}}}^2}}}+ 
{\frac{2}{{{{a_{1,2}}}^6}{{{a_{1,3}}}^2}{{{a_{2,3}}}^2}}}. 
\end{split}
\end{equation}
The Seiberg--Witten prepotential for the $A_2$ case was
also studied in \cite{KlLeTh} by using the Seiberg--Witten curve,
and the result agree with that of Nekrasov's formula
if we rescale $\Lambda^6$ to $\frac{\Lambda^6}{4}$. 
The inverse Laplace transform of each term 
is as follows. 
Recall that $a_{1,3}=a_{1,2}+a_{2,3}$.
For $\alpha,\beta,\gamma\in{\bf Z}_{\geq0}$,
\begin{equation}
\begin{split}
&{\mathcal L}_{(d_1,a_{1,2})}^{-1}\circ
{\mathcal L}_{(d_2,a_{2,3})}^{-1}\Big(
\frac{1}{{a_{1,2}}^{\alpha}{a_{2,3}}^{\beta}{a_{1,3}}^{\gamma}}
\Big)\\
&=\theta(d_2-d_1)\sum_{j=0}^{\beta-1}d_1^{\alpha+\gamma+j-1}
d_2^{\beta-1-j}
\frac{(-1)^j(\gamma)_j}{j!\Gamma(\beta-j)\Gamma(\alpha+\gamma+j)}
+(d_1,\alpha\leftrightarrow d_2,\beta)
\end{split}\label{inverselap}
\end{equation}
When $\alpha=0$ (resp. $\beta=0$), 
the second (resp. first) term is just zero.
$\theta(x)$ is the Heaviside step function.
Therefore from (\ref{swa2})(\ref{inverselap}), we obtain
\begin{equation}\label{as}
\begin{split}
r^{(0)}_1&=
(-1)^{m+1}\big(
d_2(2d_1-d_2)\theta(d_1-d_2)+d_1(2d_2-d_1)\theta(d_2-d_1)\big),
\\
r^{(0)}_2&=
-\frac{1}{2\cdot 6!}{d_2}^3
(5{d_1}^4-10{d_1}^3d_2+9{d_1}^2{d_2}^2-4d_1{d_2}^3+2{d_2}^4)
(2d_1-d_2)\theta(d_1-d_2)\\
&+(d_1\leftrightarrow d_2).
\end{split}
\end{equation}
\remark: {\bf Weyl invariance.} 
\\
The asymptotic forms 
are symmetric with respect to 
$d_1\leftrightarrow d_2$
and have the factor
$2d_1-d_2$ (resp. $2d_2-d_1$) when $d_1\geq d_2$ 
(resp. $d_2\geq d_1$).
These are the results of the 
Weyl $\times{\bf Z}_2$ invariance of 
${\mathcal F}_{d_{\rm B}}^{(g)}(a_{1,2},a_{2,3})$.
The former is due to the symmetry
$(a_{1,2},a_{2,3})\leftrightarrow (a_{2,3},a_{1,2})$,
which is the composition of the exchange
$1\leftrightarrow 3$ and the multiplication by $-1$. 
The latter is due to 
the exchange $(1,2)\mapsto(2,1)$ (resp. $(2,3)\mapsto(3,2)$);
when one sums up the terms in the same orbit of this 
action, its inverse Laplace transform turns out to have
the factor $2d_1-d_2$ (resp. $2d_2-d_1$).

\noindent
{\bf Gopakumar--Vafa invariants}

Let us define the generating function of 
the Gromov--Witten invariants
and the Gopakumar--Vafa invariants for 
given $d_{\rm B}$ by
\begin{equation}\label{gf}
\begin{split}
F^{(g)}_{d_{\rm B}}(q_1,q_2)
&:=\sum_{d_1,d_2=0}^{\infty}N_{g,d_{\rm B},d_1,d_2}
{q_1}^{d_1}{q_2}^{d_2},
\\
G^{(g)}_{d_{\rm B}}(q_1,q_2)
&:=\sum_{d_1,d_2=0}^{\infty}n^{g}_{d_{\rm B},d_1,d_2}
{q_1}^{d_1}{q_2}^{d_2}.
\end{split}
\end{equation}
One can immediately calculate $F^{(g)}_{d_{\rm B}}(q_1,q_2)$ by
simply expanding the logarithm of (11) as
\begin{equation}\label{fby}
\log {\mathcal Z}(X_{A_n}^m)|_{q=ig_s}
={\mathcal F}_{\rm GW}(X_{A_n}^m)
=\sum_{g=0}^{\infty}{g_s}^{2g-2}
\sum_{d_{\rm B}=0}^{\infty}{q_{\rm B}}^{d_{\rm B}}
F_{d_{\rm B}}^{(g)}(q_1,q_2).
\end{equation}
And one can calculate $G^{(g)}_{d_{\rm B}}(q_1,q_2)$ 
from $F^{(g)}_{d_{\rm B}}(q_1,q_2)$ as follows
(compare (\ref{gwform}) and (\ref{gvform})):
\begin{equation}
\begin{split}\label{gby}
G^{(0)}_1(q_1,q_2)&=F^{(0)}_1(q_1,q_2),\\
G^{(0)}_2(q_1,q_2)&=F^{(0)}_2(q_1,q_2)
  -\frac{1}{2^3}F^{(0)}_1({q_1}^2,{q_2}^2),\\
G^{(1)}_1(q_1,q_2)&=F^{(1)}_1(q_1,q_2)
  -\frac{1}{12}G^{(0)}_1(q_1,q_2),\\
G^{(1)}_2(q_1,q_2)&=F^{(1)}_2(q_1,q_2)
  -\frac{1}{2}F^{(1)}_1({q_1}^2,{q_2}^2)
  -\frac{1}{12}G^{(0)}_2(q_1,q_2),\\
G^{(2)}_1(q_1,q_2)&=F^{(2)}_1(q_1,q_2)
  -\frac{1}{240}G^{(0)}_1(q_1,q_2),\\
G^{(2)}_2(q_1,q_2)&=F^{(2)}_2(q_1,q_2)
  -2F^{(2)}_1({q_1}^2,{q_2}^2)
  -\frac{1}{240}G^{(0)}_2(q_1,q_2).
\end{split}
\end{equation}
Therefore from (\ref{zdbgeq})(\ref{gf})(\ref{fby})(\ref{gby}),
one can calculate the generating function of the 
Gopakumar--Vafa invariants. 
For instance, for $m=-1$,
\begin{equation}
\begin{split}\label{exu0}
&{G}_1^{(0)}=
  \frac{1+q_1+q_2-6q_1q_2+{q_1}^2q_2+q_1{q_2}^2+{q_1}^2{q_2}^2}
       {(1-q_1)^2(1-q_2)^2(1-q_1q_2)^2},
\\
&{G}_2^{(0)}=
  \frac{-2\sum_{i,j=0}^{12}C_{ij}^{(0)}{q_1}^i{q_2}^j}
  {(1-q_1)^4(1-q_2)^4(1-q_1q_2)^4
   (1-{q_1}^2)^2(1-{q_2}^2)^2(1-{q_1}^2{q_2}^2)^2}.
\end{split}
\end{equation}
The coefficients $\{C_{ij}^{(0)}\}$
\footnote{The suffix $(0)$ means $g=0$.} 
are given by
{\tiny\begin{equation*}
\begin{array}{|r|rrrrrrrrrrrrrr|}\hline
C_{ij}^{(0)}&j&0&1&2&3&4&5&6&7&8&9&10&11&12\\ \hline
i&&&&&&&&&&&&&&\\
0&&0&0&3&4&3&0&0&0&0&0&0&0&0\\
1&&0&0&-7&-13&-13&-7&0&0&0&0&0&0&0\\
2&&3&-7&0&15&-2&15&0&-7&3&0&0&0&0\\
3&&4&-13&15&54&0&0&54&15&-13&4&0&0&0\\
4&&3&-13&-2&0&-36&-74&-36&0&-2&-13&3&0&0\\
5&&0&-7&15&0&-74&26&26&-74&0&15&-7&0&0\\
6&&0&0&0&54&-36&26&192&26&-36&54&0&0&0\\
7&&0&0&-7&15&0&-74&26&26&-74&0&15&-7&0\\
8&&0&0&3&-13&-2&0&-36&-74&-36&0&-2&-13&3\\
9&&0&0&0&4&-13&15&54&0&0&54&15&-13&4\\
10&&0&0&0&0&3&-7&0&15&-2&15&0&-7&3\\
11&&0&0&0&0&0&0&0&-7&-13&-13&-7&0&0\\
12&&0&0&0&0&0&0&0&0&3&4&3&0&0\\ \hline
\end{array}
\end{equation*}}
Then the Gopakumar--Vafa invariants are 
obtained by the series expansion of $G_{d_{\rm B}}^{(g)}$:
\\
$d_{\rm B}=1$
{\tiny\[
\begin{array}{|r|r@{\ \ }rrrrrrrrrrr|}\hline
&d_2&0&1&2&3&4&5&6&7&8&9&10\\ \hline
d_1&&&&&&&&&&&&\\
0&&1&3&5&7&9&11&13&15&17&19&21\\
1&&3&4&8&12&16&20&24&28&32&36&40\\
2&&5&8&9&15&21&27&33&39&45&51&57\\
3&&7&12&15&16&24&32&40&48&56&64&72\\
4&&9&16&21&24&25&35&45&55&65&75&85\\
5&&11&20&27&32&35&36&48&60&72&84&96\\
6&&13&24&33&40&45&48&49&63&77&91&105\\
7&&15&28&39&48&55&60&63&64&80&96&112\\
8&&17&32&45&56&65&72&77&80&81&99&117\\
9&&19&36&51&64&75&84&91&96&99&100&120\\
10&&21&40&57&72&85&96&105&112&117&120&121\\ \hline
\end{array}
\]}
\noindent
$d_{\rm B}=2$
{\tiny\[
\begin{array}{|r|r@{\ }rrrrrrrrrrr|}\hline
&d_2&0&1&2&3&4&5&6&7&8&9&10\\ \hline
d_1&&&&&&&&&&&&\\
0&&0&0&-6&-32&-110&-288&-644&-1280&-2340&-4000&-6490\\
1&&0&0&-10&-70&-270&-770&-1820&-3780&-7140&-12540&-20790\\
2&&-6&-10&-32&-126&-456&-1330&-3264&-7014&-13648&-24570&-41600\\
3&&-32&-70&-126&-300&-784&-2052&-4928&-10686&-21150&-38794&-66842\\
4&&-110&-270&-456&-784&-1584&-3360&-7260&-15120&-29666&-54656&-95094\\
5&&-288&-770&-1330&-2052&-3360&-6076&-11340&-21560&-40404&-73080&-126616\\
6&&-644&-1820&-3264&-4928&-7260&-11340&-18944&-32340&-56136&-97020&-164224\\
7&&-1280&-3780&-7014&-10686&-15120&-21560&-32340&-50868&-81312
  &-131820&-213840\\
8&&-2340&-7140&-13648&-21150&-29666&-40404&-56136&-81312&-122000
  &-185328&-284914\\
9&&-4000&-12540&-24570&-38794&-54656&-73080&-97020&-131820&-185328
  &-267652&-390390\\
10&&-6490&-20790&-41600&-66842&-95094&-126616&-164224&-213840&-284914
  &-390390&-546336\\ \hline
\end{array}
\]}
We remark here that if we plug $q_i={-\beta a_{i,i+1}}(i=1,2)$
into $G_{d_{\rm B}}^{(g)}(q_1,q_2)$, the terms  with the 
lowest degree in $\beta$ reproduce the gauge theory result
$\beta^{-2(n+1)-2g+2+n}
{\mathcal F}^{(g)}_{d_{\rm B}}(a_{1,2},a_{2,3})$.
For example in $G_1^{(0)}$, the numerator is 
$\beta^2 ({a_{1,2}}^2+{a_{1,3}}^2+{a_{2,3}}^2)$,
the denominator is 
$\beta^6 {a_{1,2}}^2{a_{1,3}}^2{a_{2,3}}^2$
and 
we can check that the lowest degree part in 
$G_1^{(0)}$ is equal to $\beta^{-4} {\mathcal F}_1^{(0)}$
where ${\mathcal F}_1^{(0)}$ has been
calculated in (\ref{swa2}).

We can see the ratio
between the Gopakumar--Vafa 
invariants 
and the asymptotic form (\ref{as})
approaches 1 
when $d_1,d_2$ become large
(figure \ref{plot1}).

\section{Conclusion}
\hspace*{2.5ex}
In this article,
we have derived the asymptotic form of the 
Gopakumar--Vafa invariants of the 
toric Calabi--Yau threefold which is the 
$A_n$-fibration over ${\bf P}^1$.
The asymptotic form is,
as it turned out,
obtained as the inverse Laplace transform of the 
corresponding term in
the logarithm of Nekrasov's partition function
for instanton counting.

\section*{Acknowledgment}
\hspace*{2.5ex}
The authors would like to thank T. Eguchi 
and H. Kanno
for helpful comments.
Y. K. is grateful to S. Hosono and T. Kawai for 
valuable discussions.

A part of the numerical computation in this work
was carried out at the Yukawa Institute 
computer facility.

\appendix
\section{Asymptotic Forms: 
$\boldsymbol{A_2}$ case}
\label{data}
\begin{align*}
r^{(0)}_1&=(-1)^{m+1}\big(
-d_2(2d_1-d_2)\theta(d_1-d_2)-d_1(2d_2-d_1)\theta(d_2-d_1)
\big),\\
r^{(0)}_2&=-\frac{1}{2\cdot 6!}{d_2}^3
  (5{d_1}^4-10{d_1}^3d_2+9{d_1}^2{d_2}^2-4d_1{d_2}^3+2{d_2}^4)
  (2d_1-d_2)\theta(d_1-d_2)\\
&+(d_1\leftrightarrow d_2),\\
r^{(0)}_3&=(-1)^{m+1}\Big[
-\frac{2}{12!}{d_2}^5
  (132{d_1}^8-528{d_1}^7d_2+1012{d_1}^6{d_2}^2-1188{d_1}^5{d_2}^3
  +1045{d_1}^4{d_2}^4
\\
&\hspace{5em}-726{d_1}^3{d_2}^5+383{d_1}^2{d_2}^6-130d_1{d_2}^7
  +26{d_2}^8)(2d_1-d_2)\theta(d_1-d_2)\\
&+(d_1\leftrightarrow d_2)
\Big],\\
r^{(0)}_4&=-\frac{1}{18!}{d_2}^7
  (74919{d_1}^{12}-449514{d_1}^{11}d_2+1315171{d_1}^{10}{d_2}^2
  -2455310{d_1}^9{d_2}^3\\
&\hspace{5em}+3375333{d_1}^8{d_2}^4-3714126{d_1}^7{d_2}^5
  +3402567{d_1}^6{d_2}^6-2575908{d_1}^5{d_2}^7\\
&\hspace{5em}
  +1577430{d_1}^4{d_2}^8-756534{d_1}^3{d_2}^9+270468{d_1}^2{d_2}^{10}
  -64496d_1{d_2}^{11}\\
&\hspace{5em}+8062{d_2}^{12})(2d_1-d_2)\theta(d_1-d_2)\\
&+(d_1\leftrightarrow d_2),\\
r^{(0)}_5&=(-1)^{m+1}\Big[
-\frac{2}{5\cdot 24!}{d_2}^9
  (85968388{d_1}^{16}-687747104{d_1}^{15}d_2+2703858840{d_1}^{14}{d_2}^2\\
&\hspace{5em}
  -6891437560{d_1}^{13}{d_2}^3+12945418808{d_1}^{12}{d_2}^4
  -19376317800{d_1}^{11}{d_2}^5\\
&\hspace{5em}+24280022404{d_1}^{10}{d_2}^6
  -26073264140{d_1}^9{d_2}^7+24117002613{d_1}^8{d_2}^8\\
&\hspace{5em}
  -19125226340{d_1}^7{d_2}^9+12883936450{d_1}^6{d_2}^{10}
  -7261935912{d_1}^5{d_2}^{11}\\
&\hspace{5em}+3345086558{d_1}^4{d_2}^{12}
  -1213991290{d_1}^3{d_2}^{13}+327303495{d_1}^2{d_2}^{14}\\
&\hspace{5em}
  -58677410d_1{d_2}^{15}+5334310{d_2}^{16})(2d_1-d_2)\theta(d_1-d_2)\\
&+(d_1\leftrightarrow d_2)
\Big].
\end{align*}
\begin{align*}
r^{(1)}_1&=0,\\
r^{(1)}_2&=\frac{3}{5\cdot 9!}{d_2}^3
  (80{d_1}^6-240{d_1}^5{d_2}+363{d_1}^4{d_2}^2-326{d_1}^3{d_2}^3\\
&\hspace{4em}
   +177{d_1}^2{d_2}^4-54{d_1}{d_2}^5+18{d_2}^6)
  (2d_1-d_2)\theta(d_1-d_2)\\
&+(d_1\leftrightarrow d_2),\\
r^{(1)}_3&=(-1)^{m+1}\Big[\frac{1}{2\cdot 15!}{d_2}^5
  (93184{d_1}^{10}-465920{d_1}^9{d_2}+1151280{d_1}^8{d_2}^2
  -1809600{d_1}^7{d_2}^3\\
&\hspace{4em}
  +2026700{d_1}^6{d_2}^4-1703364{d_1}^5{d_2}^5+1127386{d_1}^4{d_2}^6
  -595192{d_1}^3{d_2}^7\\
&\hspace{4em}
  +240704{d_1}^2{d_2}^8-65178{d_1}{d_2}^9+10863{d_2}^{10})
  (2d_1-d_2)\theta(d_1-d_2)\\
&+(d_1\leftrightarrow d_2)
\Big],\\
r^{(1)}_4&=\frac{3}{21!}{d_2}^7
  (17023392{d_1}^{14}-119163744{d_1}^{13}{d_2}+412141863{d_1}^{12}{d_2}^2
  -923722506{d_1}^{11}{d_2}^3\\
&\hspace{4em}
  +1511937483{d_1}^{10}{d_2}^4-1932300342{d_1}^9{d_2}^5
  +2015543579{d_1}^8{d_2}^6\\
&\hspace{4em}
  -1760629718{d_1}^7{d_2}^7+1297849853{d_1}^6{d_2}^8
  -798184034{d_1}^5{d_2}^9\\
&\hspace{4em}
  +400904978{d_1}^4{d_2}^{10}-159563216{d_1}^3{d_2}^{11}
  +48024252{d_1}^2{d_2}^{12}\\
&\hspace{4em}
  -9861840{d_1}{d_2}^{13}+1095760{d_2}^{14})
  (2d_1-d_2)\theta(d_1-d_2)\\
&+(d_1\leftrightarrow d_2),\\
r^{(1)}_5&=(-1)^{m+1}\Big[
\frac{2}{7\cdot 26!}{d_2}^9
  (8030422400{d_1}^{18}-72273801600{d_1}^{17}{d_2}
  +321891762920{d_1}^{16}{d_2}^2\\
&\hspace{4em}
  -936927933760{d_1}^{15}{d_2}^3+2008087382840{d_1}^{14}{d_2}^4
  -3394094432680{d_1}^{13}{d_2}^5\\
&\hspace{4em}
  +4734232437620{d_1}^{12}{d_2}^6-5615707782160{d_1}^{11}{d_2}^7
  +5764569968010{d_1}^{10}{d_2}^8\\
&\hspace{4em}
  -5153404272590{d_1}^9{d_2}^9+4005220160655{d_1}^8{d_2}^{10}
  -2687880271720{d_1}^7{d_2}^{11}\\
&\hspace{4em}
  +1541655660080{d_1}^6{d_2}^{12}-744490921270{d_1}^5{d_2}^{13}
  +296060372105{d_1}^4{d_2}^{14}\\
&\hspace{4em}
  -93685509760{d_1}^3{d_2}^{15}+22307121734{d_1}^2{d_2}^{16}
  -3590362824{d_1}{d_2}^{17}\\
&\hspace{4em}
  +299196902{d_2}^{18})
  (2d_1-d_2)\theta(d_1-d_2)\\
&+(d_1\leftrightarrow d_2)\Big].
\end{align*}
\begin{align*}
r^{(2)}_1&=0,\\
r^{(2)}_2&=-\frac{1}{6\cdot 10!}{d_2}^3
  (55{d_1}^6-165{d_1}^5{d_2}
  +246{d_1}^4{d_2}^2-217{d_1}^3{d_2}^3\\
&\hspace{4em}
  +105{d_1}^2{d_2}^4-24{d_1}{d_2}^5+8{d_2}^6)
  (2d_1-d_2)({d_1}^2-d_1d_2+{d_2}^2)\theta(d_1-d_2)\\
&+(d_1\leftrightarrow d_2),\\
r^{(2)}_3&=(-1)^{m+1}\Big[
-\frac{4}{3\cdot 16!}{d_2}^5
  (19656{d_1}^{10}-98280{d_1}^9{d_2}+238524{d_1}^8{d_2}^2
  -364416{d_1}^7{d_2}^3\\
&\hspace{4em}
+388505{d_1}^6{d_2}^4
  -302835{d_1}^5{d_2}^5+182286{d_1}^4{d_2}^6
  -88439{d_1}^3{d_2}^7+33435{d_1}^2{d_2}^8\\
&\hspace{4em}
  -8436{d_1}{d_2}^9+1406{d_2}^{10})
  (2d_1-d_2)({d_1}^2-d_1d_2+{d_2}^2)\theta(d_1-d_2)\\
&+(d_1\leftrightarrow d_2)
\Big],\\
r^{(2)}_4&=-\frac{1}{2\cdot 22!}{d_2}^7
  (107343027{d_1}^{14}-751401189{d_1}^{13}{d_2}
  +2568281452{d_1}^{12}{d_2}^2\\
&\hspace{4em}
  -5641473255{d_1}^{11}{d_2}^3+8957835684{d_1}^{10}{d_2}^4
  -10984068587{d_1}^9{d_2}^5\\
&\hspace{4em}
  +10884203580{d_1}^8{d_2}^6-8988502881{d_1}^7{d_2}^7
  +6266334159{d_1}^6{d_2}^8\\
&\hspace{4em}
  -3651235662{d_1}^5{d_2}^9+1737697440{d_1}^4{d_2}^{10}
  -656195438{d_1}^3{d_2}^{11}\\
&\hspace{4em}
  +188620176{d_1}^2{d_2}^{12}-37438506{d_1}{d_2}^{13}\\
&\hspace{4em}
  +4159834{d_2}^{14})
  (2d_1-d_2)({d_1}^2-d_1d_2+{d_2}^2)\theta(d_1-d_2)\\
&+(d_1\leftrightarrow d_2),\\
r^{(2)}_5&=(-1)^{m+1}\Big[
-\frac{2}{28!}{d_2}^9
  (50517013976{d_1}^{18}-454653125784{d_1}^{17}{d_2}
  +2008600020180{d_1}^{16}{d_2}^2\\
&\hspace{3em}
  -5763329310336{d_1}^{15}{d_2}^3
  +12097712620980{d_1}^{14}{d_2}^4-19894873394844{d_2}^5{d_1}^{13}\\
&\hspace{3em}
  +26838758944032{d_2}^6{d_1}^{12}-30659562087156{d_2}^7{d_1}^{11}
  +30245624065647{d_2}^8{d_1}^{10}\\
&\hspace{3em}
  -25967123885015{d_2}^9{d_1}^9
  +19370347312179{d_2}^{10}{d_1}^8-12464676224742{d_2}^{11}{d_1}^7\\
&\hspace{3em}
  +6849482740752{d_2}^{12}{d_1}^6-3169505142660{d_2}^{13}{d_1}^5
  +1209745627491{d_2}^{14}{d_1}^4\\
&\hspace{3em}
  -368804218845{d_2}^{15}{d_1}^3
  +85189452681{d_2}^{16}{d_1}^2-13450408536{d_2}^{17}{d_1}\\
&\hspace{3em}
  +1120867378{d_2}^{18})
  (2d_1-d_2)({d_1}^2-d_1d_2+{d_2}^2)\theta(d_1-d_2)\\
&+(d_1\leftrightarrow d_2)
\Big].
\end{align*}

\section{Local B-model Calculation: $\boldsymbol{A_2}$ case} 
\label{localB}
\hspace*{2.5ex}
We computed the Gopakumar--Vafa invariants
of $A_2$-fibration over ${\bf P}^1$,
for $g=0$ and $d_{\rm B}\leq 2$, $d_1,d_2\leq 21$
by the local B-model calculation \cite{ChKlYaZa}.
The results agree with the results from
the partition function of Iqbal and Kashani-Poor
(\ref{zdbgeq}) and the results 
in \cite{ChKlYaZa} (section 6.4) for
$m=-1$ and $m=2$.
In this section, we list the relevant data.

\noindent ``Charge vectors'':
\begin{align*}
\begin{bmatrix}
l^{(0)}\\l^{(1)}\\l^{(2)}
\end{bmatrix}
&=\begin{cases}
\begin{bmatrix}
1&1&-m&-2+m&0&0\\0&0&1&-2&1&0\\0&0&0&1&-2&1
\end{bmatrix}
&(m=0,1,2)\\
\begin{bmatrix}
1&1&0&-1&-1&0\\0&0&1&-2&1&0\\0&0&0&1&-2&1
\end{bmatrix}
&(m=-1)
\end{cases}
\end{align*}
Gr\"obner basis of the toric ideal $I_A$:
\begin{equation*}
l^{(0)},\quad l^{(1)},\quad l^{(2)},\quad l^{(1)}+l^{(2)}.
\end{equation*}
Solutions to
GKZ-system $H_A(\beta)$ with $\beta=(0,0,0)$:
\begin{equation}\label{solutions}
f,\quad
-t_B=\partial_{\rho_0}f,\quad -t_1=\partial_{\rho_1}f,
\quad-t_2=\partial_{\rho_2}f,
\quad I_1,I_2.
\end{equation}
with
\begin{align*}
&m=-1:&&
I_1=(2\partial_{\rho_0}\partial_{\rho_1}+\partial_{\rho_1}^2)f,
&&
I_2=(2\partial_{\rho_0}\partial_{\rho_2}+\partial_{\rho_2}^2)f.
\\
&m=0:&&
I_1=\partial_{\rho_0}\partial_{\rho_1}f,
&&
I_2=(\partial_{\rho_0}\partial_{\rho_2}+\partial_{\rho_2}^2)f.
\\
&m=1:&&
I_1=(2\partial_{\rho_0}\partial_{\rho_1}+\partial_{\rho_1}^2)f,
&&
I_2=(2\partial_{\rho_2}\partial_{\rho_0}+2\partial_{\rho_1}
\partial_{\rho_2}+3\partial_{\rho_2}^2)f.
\\
&m=2:&&
I_1=(\partial_{\rho_0}\partial_{\rho_1}+\partial_{\rho_1}^2)f,
&&
I_2=(\partial_{\rho_0}\partial_{\rho_2}+
2\partial_{\rho_1}\partial_{\rho_2}+2\partial_{\rho_2}^2)f.
\end{align*}
Here
\begin{equation*}
f=\sum_{n_0,n_1,n_2\geq 0}
\prod_{i=1}^{6}\frac{
\Gamma\big(\sum_{j=0}^2\rho_j(l^{(j)})_i+1\big)}
{\Gamma\big(\sum_{j=0}^2(\rho_j+n_j)(l^{(j)})_i+1)\big)}
\cdot
{z_0}^{n_0+\rho_0}{z_1}^{n_1+\rho_1}
{z_2}^{n_2+\rho_2}
\end{equation*}
Identification with the prepotential (here,
${\mathcal F}={\mathcal F}_{\rm GW}^{g=0}(X_{A_n}^m)$):
\begin{align*}
&m=-1:
&&I_1=2(-\partial_{t_B}-2\partial_{t_1}+\partial_{t_2})
{\mathcal F}&&
I_2=2(-\partial_{t_B}+\partial_{t_1}-2\partial_{t_2}){\mathcal F}
\\
&m=0:
&&I_1=(-2\partial_{t_B}-2\partial_{t_1}+\partial_{t_2}){\mathcal F}&&
I_2=(\partial_{t_1}-2\partial_{t_2}){\mathcal F}
\\
&m=1:
&&I_1=2(-\partial_{t_B}-2\partial_{t_1}+\partial_{t_2}){\mathcal F}&&
I_2=2(\partial_{t_1}-2\partial_{t_2}){\mathcal F}
\\
&m=2:&&
I_1=(-2\partial_{t_1}+\partial_{t_2}){\mathcal F}&&
I_2=(\partial_{t_1}-2\partial_{t_2}){\mathcal F}
\end{align*}
We have determined the overall  normalization 
so that $n_{1,1,0}=n_{1,0,1}=n_{1,1,1}=-2$.


\end{document}